\documentclass[aps,amsfonts,superscriptaddress, nofootinbib, floatfix, 10pt,twocolumn,prd]{revtex4-1} 
\usepackage{graphicx}
\usepackage{amsmath}
\usepackage[latin1]{inputenc}
\usepackage{amsbsy}
\usepackage{color}
\usepackage{epsfig}
\usepackage{graphicx}
\usepackage{amsmath}
\usepackage{caption}
\usepackage{subcaption}
\usepackage{amssymb}
\usepackage{bbm}
\usepackage{amsbsy}
\usepackage{slashed}

\usepackage{xcolor}

\usepackage{soul}

\usepackage[normalem]{ulem}  

\def\be{\begin{equation}}

\def\ee{\end{equation}}
\def\bea{\begin{eqnarray}}
\def\eea{\end{eqnarray}}

\begin{document} 


\title{A new perspective for the magnetic corrections to  $\pi$-$\pi$ Scattering Lengths in the Linear Sigma Model}

\author{M.~Loewe}
\email{mloewe@fis.puc.cl}
\affiliation{Instituto de F\'isica, Pontificia Universidad Cat\'olica de Chile, Casilla 306, San\-tia\-go 22, Chile}
\affiliation{Centre for Theoretical and Mathematical Physics and Department of Physics,
University of Cape Town, Rondebosch 7700, South Africa}
\affiliation{Centro Cient\'ifico Tecnol\'ogico de Valpara\'iso-CCTVAL, Universidad T\'ecnica Federico Santa Mar\'ia, Casilla 110-V, Valpara\'iso, Chile} 
\author{L. Monje}
\email{lnmonje@uc.cl}
\affiliation{Instituto de F\'isica, Pontificia Universidad Cat\'olica de Chile, Casilla 306, San\-tia\-go 22, Chile}
\author{E. Mu\~noz}
\email{munozt@fis.puc.cl}
\affiliation{Instituto de F\'isica, Pontificia Universidad Cat\'olica de Chile, Casilla 306, San\-tia\-go 22, Chile}
\author{A. Raya}
\email{raya@ifm.umich.mx}
\affiliation{Instituto de F\'isica y Matem\'aticas, Universidad Michoacana de San Nicol\'as de Hidalgo, Edificio C-3, Ciudad Universitaria, Morelia 58040, Michoac\'an, M\'exico.}
\author{R. Zamora}
\email{rzamorajofre@gmail.com}
\affiliation{Instituto de Ciencias B\'asicas, Universidad Diego Portales, Casilla 298-V, Santiago, Chile} 
\affiliation{Centro de Investigaci\'on y Desarrollo de Ciencias Aeroespaciales (CIDCA), Fuerza A\'erea de Chile, Santiago 8020744, Chile}



\begin{abstract}
In this article, a new perspective for obtaining the magnetic evolution of $\pi-\pi $ scattering lengths in the frame of the linear sigma model is presented.  When computing the relevant one-loop diagrams that contribute to these parameters, the sum over Landau levels --emerging from the expansion of the Schwinger propagator-- is handled in a novel way that could also be applied to the calculation of other magnetic-type corrections.  Essentially, we have obtained an expansion in terms of Hurwitz Zeta functions. It is necessary to regularize our expressions by an appropriate physical subtraction when $|qB| \rightarrow 0$ ($q$ the meson charge and $B$ the magnetic field strength). In this way, we are able to interpolate between the very high magnetic field strength region, usually handled in terms of the Lowest Landau Level (LLA) approximation, and the weak field region, discussed in a previous paper by some of us, which is based on an appropriate expansion of the Schwinger propagator up to order $|qB|^{2}$. Our results for the scattering lengths parameters produce a soft evolution in a wide region of magnetic field strengths, reducing to the previously found expressions in both limits.

\end{abstract}


\maketitle


\section{Introduction}
During the last years,  physics of strongly interacting hadron matter at high temperatures and densities, or baryonic chemical potential, including huge magnetic fields generated during peripheral heavy ion collisions, has attracted the attention of the community, both from the experimental as well as from the theoretical point of view. Different  experiments have reported  interesting results that are related to temperature and eventually to magnetic corrections of some physical quantities. For example, an excess of photons at low momentum in the invariant momentum distribution has been reported~\cite{ref1,ref2,ref3}. It has been argued~\cite{photons} that this excess, after taking into account common sources like synchroton radiation, Bremsstrahlung or pair-annihilation, could be related to gluon fusion at early times of the collision. This mechanism  is only possible if magnetic effects are present. Signals related only to temperature effects are much more well established as, for example, the broadening of hadron resonances~\cite{ref4}. Of course, the effect on broadening of resonances widths should take into account simultaneous corrections in temperature and magnetic field.
 
New experiments, like NICA~\cite{nica}, would be able to explore density effects, allowing the experimental discussion of scenarios like quarkyonic matter~\cite{ref5}. Of course, a dense nuclear matter environment is one of the crucial ingredients for our understanding of compact objects like neutron stars. Lattice groups~\cite{ref6} have also found the remarkable phenomenon of inverse magnetic catalysis, which corresponds  to the decreasing of both the pseudocritical  temperature  for the chiral and/or deconfinement phase transitions, and for the quark condensates, as function of an increasing magnetic field strength. Different explanations have been proposed, as for example, an analysis that goes beyond the mean field approximation, by considering thermomagnetic corrections to the couplings as well as plasma screening for the boson masses through ring diagrams, in the linear sigma model \cite{ref7}. It is worthwhile to mention that, as a consequence of a renormalization group analysis, an analytical expression for the thermo-magnetic evolution of the QCD strong coupling has recently been found \cite{ref8}. This should allow to extend existing theoretical calculations.

It seems, therefore,  that the discussion of the magnetic dependence of different physical parameters  is highly relevant for our understanding of these kind of physical phenomena.  In general, it is not an easy task to disentangle magnetic effects from other kinds of corrections, and this represents an important motivation to search for new channels or parameters where this goal could be achieved.  $\pi$~-~$\pi $ scattering lengths are responsible for the interaction among pions at low energies, near the threshold. Since  about 600 pions are produced in a single heavy ion collision event, and since charmonium and bottomonium states --that may survive beyond the critical temperature-- decay into pairs of pions whose re-scattering has been measured, we see that scattering lengths could be relevant parameters for a better understanding of the magnetic and thermalf evolution. In fact, the so called cusp-effect in the emerging pions from such heavy onium states  has been used as a clear signal for measuring $\pi $-$\pi$ scattering lengths~\cite{ref9}. It is certainly a challenge to measure such signals in heavy ion collision experiments. The new results we are presenting here are not restricted anymore to the low magnetic field regime, as was the case in a previous article by some of us~\cite{Leandro}.
 
Some years ago, analysis were done on the temperature dependence of these scattering lengths parameters using the Nambu-Jona-Lasinio~\cite{NJL} and the linear sigma~\cite{TLSM} models. As previously mentioned, magnetic effects on these objects were computed in the linear sigma model by some of us using an expansion of the Schwinger propagator valid for small magnetic fields~\cite{Leandro}. The main result of that article points out the opposite effect of the magnetic field and temperature was interesting, since it seems that magnetic and temperature effects are opposite to each other. For the isospin $I=0,2$ channels, the $\pi $-$\pi $ scattering lengths turn out to increase/decrease as a function of temperature. The opposite effects were found for the magnetic evolution.

Here we present, in the linear sigma model at the one-loop level, a new discussion on the magnetic dependence of the $\pi $-$\pi $ scattering lengths, valid for arbitrary values of the magnetic field strength. The novelty of the analysis relies on the way we handle the relevant integrals that appear in the one-loop diagrams. In fact, using the well known expansion for the Schwinger propagator in terms of Landau levels, and introducing a physically transparent regularization of a certain magnetic field dependent logarithmic divergent term, we are able, as we present in the next sections, to obtain quite compact expansions for the relevant one loop integrals associated to the $s$- and $t$-channel contributions. The remaining of the paper is organized as follows: In section II the linear sigma model is revised, presenting the $\pi $- $\pi $ scattering lengths decomposed according to isospin channel projections. Then, in section III the detailed computation of the magnetic field contribution to the $\pi $- $\pi $ scattering lengths is presented, including the regularization of a magnetic-dependent divergent term. In this way, we are able to present our results for the magnetic evolution for the scattering lengths in both relevant isospin channels $I = 0, 2$. More technical details are presented in an appendix. Finally, in section IV we present our final conclusions.

\section{Linear sigma model and $\pi$-$\pi$ scattering}

The linear sigma model was introduced by Gell-Mann and L\'evy~\cite{Gell-Mann} as an effective approach to describe chiral symmetry breaking via an explicit and spontaneous mechanism.  In the context of critical phenomena, the model represents a field theory where the Lagrangian possesses $O(N)$ symmetry, which near the critical temperature is spontaneously broken into $O(N-1)$, thus
leading to $N-1$ massless Goldstone bosons (representing tangential oscillating modes), and a single massive field (representing radial oscillations) with respect to the minimum of a mexican-hat shaped effective potential.

In the phase where the chiral symmetry is
broken, the model is given by  
\begin{eqnarray}
\mathcal{L}&=&\bar{\psi}\left[i\gamma^{\mu}\partial_{\mu}-m_{\psi}-g(s+i\vec{\pi}\cdot\vec{\tau}\gamma_{5})\right]\psi\\
&&+\frac{1}{2}\left[(\partial\vec{\pi})^2+m_{\pi}^2\vec{\pi}^2\right]+\frac{1}{2}\left[(\partial\sigma)^2+m_{\sigma}^2 s^2\right]\nonumber\\
&&-\lambda^2vs(s^2+\vec{\pi}^2)-\frac{\lambda^2}{4}(s^2+\vec{\pi}^2)^2+(\varepsilon c-vm_{\pi}^2)s.\nonumber
\end{eqnarray}
In this expression $v=\langle\sigma\rangle$ is the vacuum expectation value of the
scalar field $\sigma$. The idea is to define a new field $s$ such
that $\sigma = s+v$, with $\langle s\rangle=0$. $\psi$
corresponds to an isospin doublet associated to the nucleons,
$\vec{\pi}$ denotes the pion isotriplet field and $c\sigma$ is the
term that breaks explicitly the $SU(2)\times SU(2)$ chiral
symmetry. $\epsilon$ is a small dimensionless parameter. It is
interesting to remark that all fields in the model have masses
determined by $v$. In fact, the following relations are valid:
$m_{\psi}=gv$, $m_{\pi}^2=\mu^2+\lambda^2v^2$ and
$m_{\sigma}^2=\mu^2+3\lambda^2v^2$. Perturbation theory at the
tree level allows us to identify the pion decay constants as
$f_{\pi}=v$. Finite temperature effects on this model have been studied by several authors, discussing the thermal
evolution of masses, $f_\pi(T)$, the effective potential, etc.~\cite{Loewe,Larsen,Bilic,Petropolus,wagner,kovacs1,kovacs2,kovacs3}.

Since our idea is to use the linear sigma model for calculating
$\pi$-$\pi$ scattering lengths, let us remind briefly the
formalism. A scattering amplitude has the general form~\cite{Collins, Gasser}

\begin{eqnarray}
T_{\alpha\beta;\delta\gamma}&=&A(s,t,u)\delta_{\alpha\beta}\delta_{\delta\gamma}+A(t,s,u)\delta_{\alpha\gamma}\delta_{\beta\delta}\nonumber\\
&&+A(u,t,s)\delta_{\alpha\delta}\delta_{\beta\gamma},
\label{proyectores}
\end{eqnarray}
\noindent where $\alpha$, $\beta$, $\gamma$, $\delta$ denote
isospin components.

By using appropriate projection operators, it is possible to find
the following isospin dependent scattering amplitudes



\begin{align}
T^{0}&=3A(s,t,u)+A(t,s,u)+A(u,t,s),\label{eq3}\\
T^{1}&=A(t,s,u)-A(u,t,s),\label{eq4}\\
T^{2}&=A(t,s,u)+A(u,t,s),
\label{eq5}
\end{align}

\noindent where $T^I$ denotes a scattering amplitude in a given isospin channel $I = \{0,1,2\}$.\\

As it is well known~\cite{Collins}, the isospin dependent
scattering amplitude can be expanded in partial waves $T_\ell^I$,

\begin{equation}
T_{\ell}^{I}(s)=\frac{1}{64\pi}\int_{-1}^{1}d(cos\theta)P_{\ell}(cos\theta)T^{I}(s,t,u).
\end{equation}

Below the inelastic threshold, the partial scattering amplitudes
can be parametrized as~\cite{Gasser}
\begin{equation}
T_{\ell}^{I}=\left(\frac{s}{s-4m\pi^2}\right)^{\frac{1}{2}}\frac{1}{2i}\left(e^{2i\delta_{\ell}^{I}(s)}-1\right),
\end{equation}

\noindent where $\delta_{\ell}$ is a phase-shift in the $\ell$
channel. The scattering lengths are important parameters in order
to describe low energy interactions. In fact, our last expression
can be expanded according to
\begin{equation}
\Re\left(T_{\ell}^{I}\right)=\left(\frac{p^{2}}{m_{\pi}^{2}}\right)^{\ell}\left(a_{\ell}^{I}+\frac{p^2}{m_{\pi}^{2}}b_{\ell}^{I}+\ldots\right).
\end{equation}
The parameters $a_{\ell}^{I}$ and $b_{\ell}^{I}$ are the
scattering lengths and scattering slopes, respectively. In
general, the scattering lengths obey $|a_{0}^{I}|>|a_{1}^{I}|>|a_{2}^{I}|...$.
If we are only interested in the scattering lengths $a_0^I$, it is
enough to calculate the scattering amplitude $T^I$ in the static
limit, i.e. when $s \to 4m_\pi^2$, $t\to 0$ and $u\to 0$
\begin{equation}
a_{0}^{I}=\frac{1}{32\pi}T^{I}\left(s \to 4m_{\pi}^2,t\to 0, u\to0\right).\label{eq:a0I}
\end{equation} 
 The first measurement of $\pi $-$\pi $ scattering lengths was carried on by Rosellet 
{\em et al.}
~\cite{Rosellet}. More recently, these parameters have been measured  using pionium atoms in the DIRAC experiment \cite{DIRAC} and also through the decay of heavy quarkonium states into $\pi $-$\pi $ final states where the so called cusp-effect was found~\cite{quarkonium}. We evaluate  expression~(\ref{eq:a0I}) for $I=0,2$ in a background magnetic field of arbitrary strength below.

\section{Scattering lengths at finite magnetic field}
Recently, some of us discussed the magnetic evolution of the $\pi$-$\pi$ scattering lengths in the frame of the linear sigma model~\cite{Leandro} 

Our analysis was based on a perturbative treatment of the bosonic Schwinger propagator, valid for small magnetic fields. We found that this magnetic evolution displays an opposite trend with respect to thermal corrections on the scattering lengths, reported previously in the literature \cite{TLSM}. At low magnetic field intensities, the scattering lengths in the isospin channel $I = 2$ increase whereas their projection into the channel $I = 0$ diminishes, both as function of the magnetic field. It is interesting to re-analyze this problem in the full range of magnetic field intensities. In fact, in  peripheral heavy ion collisions we may expect extremely high magnetic fields, that may affect the interaction among the emerging pions generated during the collision.

\bigskip

\begin{figure}[h!]
{\centering
\includegraphics[scale=0.75]{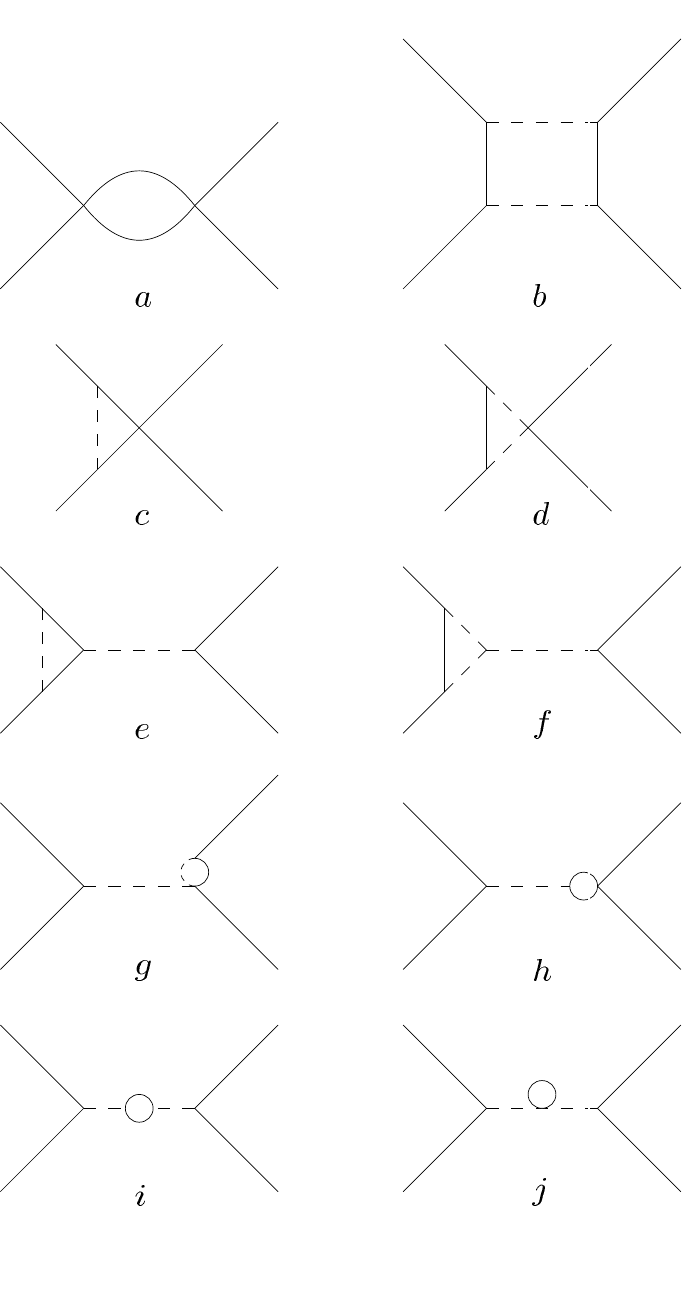}}
\caption{One-loop diagrams relevant to the $\pi$-$\pi$ scattering lengths}. Continuous and dashed lines represent pions and $\sigma$ mesons respectively.
\label{oneloop}
\end{figure}
\noindent In the linear sigma model, the relevant diagrams that contribute to $\pi$-$\pi$ scattering are shown in Fig. \ref{oneloop}. Notice that tadpole-like diagrams associated to mass corrections of the sigma field, do not contribute to the  $\pi$-$\pi$ scattering amplitudes, because they do not possess an absortive component, 
since their imaginary part is zero. These tadpoles are extremely small in the limit of a very large mass of the sigma field. This approximation is valid since, as we know, $m_{\sigma} \approx 550$~MeV is much larger than the pion mass. Fermions, i.e. nucleons that may interact with pions, are not considered in our discussion. As a consequence, the sigma field propagator is contracted to a point.

\bigskip
\begin{figure}
        \begin{subfigure}[b]{0.5\textwidth}
		\centering
                \includegraphics[scale=0.6]{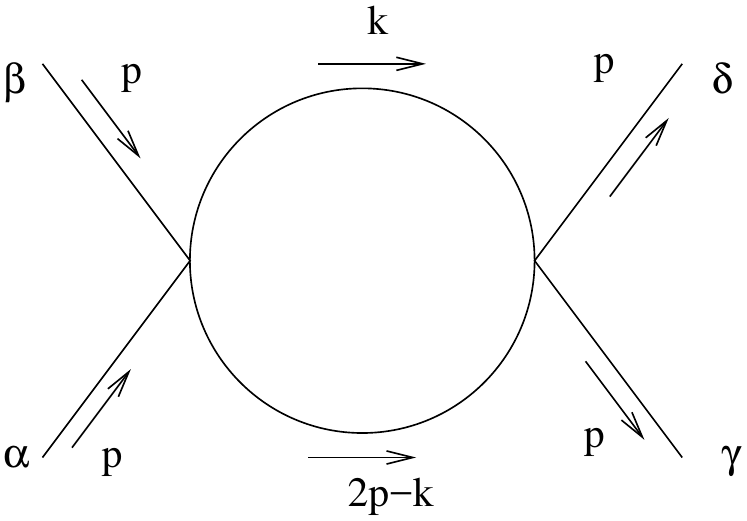}
                \caption{$s$-channel diagram.}
                \label{fig:gull}
        \end{subfigure}
        \begin{subfigure}[b]{0.5\textwidth}
		\centering
                \includegraphics[scale=0.6]{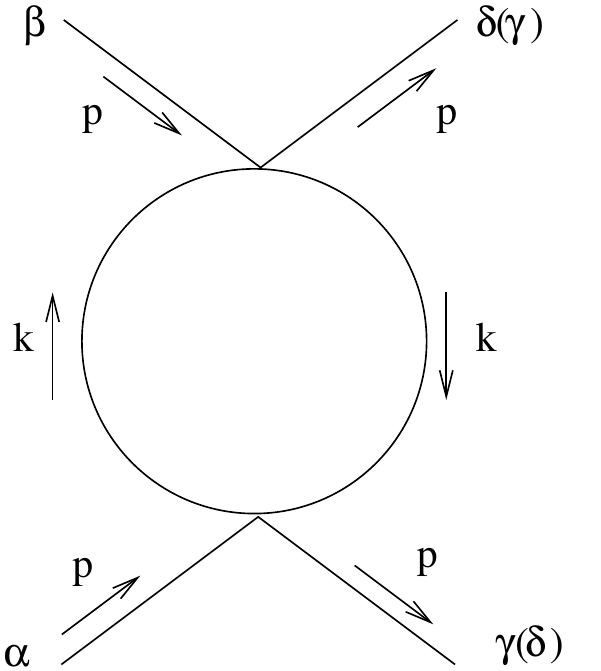}
                \caption{$t$ and $u$ channel diagram.}
                \label{fig:gull2}
        \end{subfigure}
        \caption{``Fish-type'' diagrams.}\label{fig:a diagrams}
\end{figure}

\noindent From these considerations, we see that all relevant diagrams reduce to a horizontal ($s$-channel) or vertical ($t$ and $u$ channels) {\em``fish-type"} pion loops contributions, as shown in Fig.~\ref{fig:a diagrams}. Then, our task is to compute such diagrams as a function of the magnetic field intensity. This is an interesting problem, not only because of physical implications, but also due to new analytical results that we present below.

\bigskip
\noindent Let us derive our starting expression for the bosonic propagator as a sum of Landau levels. The bosonic Schwinger propagator for a charged pion of charge $q$ immersed in a uniform magnetic field along the third spatial coordinate, in the proper time representation is given by
\begin{align}
iD^{B} (k)&=\int_0^\infty \frac{ds}{\cos(qBs)}e^{ is\left(k_{||}^2-k_\perp^2\frac{\tan(qBs)}{qBs}-m_\pi^2 +i\epsilon \right)}.
\end{align}
 
\noindent After inserting this propagator in the  fish-type diagrams, it is not difficult to see that all contributions reduce to two types of integrals
\begin{eqnarray}
I_1[B,p_0]	&=&\int\frac{d^4k}{(2\pi)^4}iD^{B}(k_0,\mathbf k)iD^{B}(k_0-2p_0,\mathbf k), \label{int1}\\
I_2[B]	&=& \int\frac{d^4k}{(2\pi)^4} \left[iD^{B}(k_0,\mathbf k)\right]^2 \label{int2}.
\label{eq_I12}
\end{eqnarray}

For technical purposes, we shall calculate the integrals with the expression for the propagator at finite magnetic field in terms of Landau levels, as presented in \cite{Ayalaetal}.
\begin{eqnarray}
i D^{B}(k) = 2  \sum_{l=0}^{\infty}(-1)^{l} L_{l}\left(\frac{2 k_{\perp}^2}{qB} \right) e^{-k_{\perp}^2/qB} i\Delta_{l}^{B}(k_{\parallel}),
\label{eq_prop}
\end{eqnarray}
where $L_{l}(z)$ are the Laguerre polynomials, and 
we have defined the effective ``parallel'' propagators
\begin{eqnarray}
i\Delta_{l}^{B}(k_{\parallel}) = \frac{i}{k_{\parallel}^{2} - (2l + 1)qB - m_{\pi}^2 + i\epsilon }.
\label{eq_prop_Land}
\end{eqnarray}

Let us first consider the calculation of $I _{1}[B]$, after its definition in Eq.(\ref{eq_I12}), substituting the infinite series for the propagators, Eq.(\ref{eq_prop}), we are lead to
\begin{eqnarray}
I_1[B,p_0] &&= \int \frac{d^2 k_{\parallel}\,d^2 k_{\perp}}{(2\pi)^4} i D^{B}(k) iD^{B}(k_0 - 2p_0,\mathbf{k})\nonumber \\
&& = 4 \sum_{l=0}^{\infty} \sum_{l' = 0}^{\infty}(-1)^{l + l'} 
G_{l,l'}(p_0)\\
&&\times \left[\int\frac{d^2 k_{\perp}}{(2\pi)^2} e^{-2\,k_{\perp}^2/qB}  L_{l}\left(\frac{2 k_{\perp}^2}{qB} \right) L_{l'}\left(\frac{2 k_{\perp}^2}{qB} \right)\right].\nonumber
\label{eq_double_sum}
\end{eqnarray}
Here, we have defined the integrals
\begin{eqnarray}
G_{l,l'}(p_0) = \int \frac{d^2 k_{\parallel}}{(2\pi)^2} i\Delta_{l}^{B}(k_{\parallel}) i\Delta_{l'}^{B}(k_{\parallel} - 2p_0). 
\end{eqnarray}

Let us now calculate the integral over the Laguerre polynomials in the second term, by using 2-dimensional ``spherical coordinates",
with $0 \le | k_{\perp}| < \infty$,
\begin{eqnarray}
d^{2}k_{\perp} &=& 2	\pi |k_{\perp}| d| k_{\perp}| = \frac{\pi\,q\,B}{2} dx,
\end{eqnarray}
where we have defined the auxiliary variable $x = 2 k_{\perp}^2/qB$, with $0 \le x < \infty$. Therefore, we have
\begin{eqnarray}
&&\int\frac{d^2 k_{\perp}}{(2\pi)^2} e^{-2\,k_{\perp}^2/qB}  L_{l}\left(\frac{2 k_{\perp}^2}{qB} \right) L_{l'}\left(\frac{2 k_{\perp}^2}{qB} \right) \nonumber \\
&=& \frac{1}{4\pi^2}\frac{\pi\,q B}{2} \int_{0}^{\infty} dx e^{-x} L_{l}(x) L_{l'}(x)\nonumber\\
&=& \frac{qB}{8\pi}\delta_{l,l'},
\label{eq_Laguerre_orto}
\end{eqnarray}
where the orthogonality relation between Laguerre polynomials was used.
Substituting this result into Eq.(\ref{eq_double_sum}), we end up with the expression
\begin{eqnarray}
I_1[B,p_0] = \frac{qB}{2\pi}\sum_{l=0}^{\infty} G_{l,l}(p_0).
\label{eq_I1_reduced}
\end{eqnarray}

As shown in detail in Appendix, we calculate $G_{l,l}(p_0)$ by first integrating over $k_0$ in the complex plane, and later over $k_3$. This procedure allows us to obtain the infinite series
\begin{eqnarray}
I_1[B,p_0] = \frac{i}{16\pi^2 }\frac{2qB}{p_0^2}
\sum_{l=0}^{\infty} z_l{\rm{Arctan}}(z_l ),
\label{eq_atan}
\end{eqnarray}
where we have defined $z_l = (p_0/\sqrt{2qB})/\sqrt{l+1/2 + (m_\pi^2 - p_0^2)/(2qB)}$. This infinite series, as expected, displays a mild logarithmic divergence, that can however be removed with a straightforward procedure, as we now show. Let us first expand
each term on the series above, using the infinite series (valid for $|z_l|>1$ and $|z_l|< 1$)
\begin{eqnarray}
z_l {\rm{Arctan}}(z_l ) &=& \sum_{m=0}^{\infty} \frac{2^{2m}(m!)^2}{(2m+1)!} \left(\frac{z_l^2}{1 + z_l^2}\right)^{m+1}\nonumber\\
&=& \sum_{m=0}^{\infty} \frac{2^{2m}(m!)^2}{(2m+1)!} \left(1 + z_l^{-2}\right)^{-m-1}.
\label{eq_atan_series}
\end{eqnarray}
Inserting Eq.(\ref{eq_atan_series}) back into Eq.(\ref{eq_atan}), and exchanging the order of
the sums, we obtain
\begin{eqnarray}
I_1[B,p_0] &=& \frac{i}{16\pi^2}\sum_{m=0}^{\infty}\frac{2^{2m}(m!)^2}{(2m+1)!}\left(\frac{p_0^2}{2qB}\right)^m\nonumber\\
&&\times\zeta\left(1 + m, \frac{1}{2} + \frac{m_{\pi}^2}{2qB}\right),
\label{eq_I1_final}
\end{eqnarray}
where $\zeta(\alpha,z) = \sum_{l=0}^{\infty}(z + l)^{-\alpha}$ are the Hurwitz Zeta functions. It is important to remark that the term $m=0$ needs to be regularized, using the relation between the Hurwitz Zeta function and the digamma function $\psi(z)$,
\begin{eqnarray}
\zeta(z,1+\epsilon) = -\psi(z) + \frac{1}{\epsilon} + O(\epsilon),\,\,\epsilon\rightarrow 0^{+}.
\end{eqnarray}
The asymptotic behavior of the digamma function for very large values of its argument ($|z| \gg 1$) is captured by the series
\begin{eqnarray}
\psi(z) \sim \ln(z) - \sum_{n=1}^{\infty}\frac{B_n}{n}\,z^{-n},
\end{eqnarray} 
where $B_k$ are the Bernouilli numbers, for $B_1 = 1/2$. Clearly, the digamma function displays a logarithmic divergence in this limit. Therefore, the expression for $I_1[B]$ in Eq.(\ref{eq_I1_final}) diverges as $B\rightarrow 0$, as expected from the vacuum contribution to the diagram at zero field. Since
we are interested in the contribution due to the finite magnetic field with respect to the experimental zero-field value of the scattering length, we define the regularized expression
\begin{eqnarray}
I_1^{Reg}[B,p_0] &&\equiv I_1[B,p_0] - I_1[B\rightarrow 0,p_0]\nonumber\\
&&= \int\frac{d^4k}{(2\pi)^4}\bigl[i D^{B}(k_0,\mathbf k)i D^{B}(k_0-2p_0,\mathbf k) \nonumber \\
&&- i D^{0}(k_0,\mathbf k)i D^{0}(k_0-2p_0,\mathbf k)\bigr],
\end{eqnarray}
where clearly, by definition
\begin{eqnarray}
\lim_{B\rightarrow 0} I_1^{Reg}[B,p_0] = 0.
\label{eq_I1_limit}
\end{eqnarray}
In order to construct the regularized form, we subtract the asymptotic, logarithmically divergent expression for the digamma function ($m=0$) at small magnetic field, as follows
\begin{eqnarray}
&&I_1^{Reg}[B,p_0] = \frac{i}{16\pi^2}\left[-\psi\left(\frac{1}{2} + \frac{m_{\pi}^2}{2 q B}\right) + \ln\left(\frac{1}{2} + \frac{m_{\pi}^2}{2 q B}\right)\right. \nonumber\\   
&&\left. +\sum_{m=1}^{\infty}\frac{(m!)^2}{(2m+1)!}\left(2\frac{p_0^2}{qB}\right)^m\,\zeta\left(m+1,\frac{1}{2} + \frac{m_{\pi}^2}{2 q B}\right)\right].
\label{eq_I1_Reg}
\end{eqnarray}

Let us now turn our attention to the integral $I_2[B]$ defined in Eq.(\ref{eq_I12}). It is straightforward to obtain the regularized
expression of this integral by setting $p_0 = 0$ as follows
\begin{eqnarray}
I_2^{Reg}[B] &&\equiv I_2[B] - I_2[B\rightarrow 0]\\ 
&=& \lim_{p_0\rightarrow 0} I_{1}^{Reg}[B,p_0]\nonumber\\
&=& \frac{i}{16\pi^2}\left[
-\psi\left(\frac{1}{2} + \frac{m_{\pi}^2}{2 q B}\right) + \ln\left(\frac{1}{2} + \frac{m_{\pi}^2}{2 q B}\right)
\right].\nonumber
\label{eq_I2_Reg}
\end{eqnarray}

\noindent In order to obtain the scattering lengths $a_0^I$, we use the decomposition of the scattering amplitude in the different isospin channels presented in Section I. 
Since we are only interested in the scattering lengths $a_0^I$, it is enough to calculate the scattering amplitude in the static limit. Therefore, we normalize by the experimental values at tree level~\cite{peyaud},  to obtain the expressions
\begin{eqnarray}
a_0^0(B)	&=&a_0^0(\text{exp})+\frac1{32\pi}\bigl(3A(s,t,u)+A(t,s,u)\nonumber \\
&+& A(u,t,s)\bigr),\nonumber\\
a_0^2(B)	&=&a_0^2(\text{exp})+\frac1{32\pi}\left(A(t,s,u)+A(u,t,s)\right).
\end{eqnarray}

Here, $A(s,t,u)$, $A(t,s,u)$ and $A(u,t,s)$ correspond to all $s$-channel, $t$-channel and $u$-channel contributions, respectively.
On the other hand, the $s$-channel contribution is obtained from $I_1^{Reg}[B,p_0=m_{\pi}]$, while those for the $t$- and $u$-channels are obtained from $I_2^{Reg}[B]$, according to the following expressions

\begin{eqnarray}
A(s,t,u) &=& -4\lambda^4\left(1 - \frac{12 \lambda^2 v^2 }{m_{\sigma}^2}+\frac{24 \lambda ^4 v^4 }{m_{\sigma }^4}\right) \nonumber \\
&\times& I_1^{Reg}[B,p_0 = m_{\pi}],\nonumber\\
A(t,s,u) + A(u,t,s) &=& -8 \lambda^4\left(
1 - \frac{12 \lambda^2 v^2}{m_{\sigma}^2} + \frac{24 \lambda ^4 v^4 }{m_{\sigma }^4}
\right) \nonumber \\
&&\times I_2^{Reg}[B].
\end{eqnarray}

The experimental values in the absence of magnetic field $B = 0$ are given by \cite{peyaud}  $a_0^0(\text{exp})=0.217$ and $a_0^2(\text{exp})=-0.041$. The mass for the sigma meson
is set to  $m_{\sigma} = 550$ MeV, and the mass for the pion $m_{\pi} = 140$ MeV, with the parameter $v = 89$ and $\lambda^2=4.26$.

\begin{figure}[t!]
\centering
\includegraphics[scale=0.3]{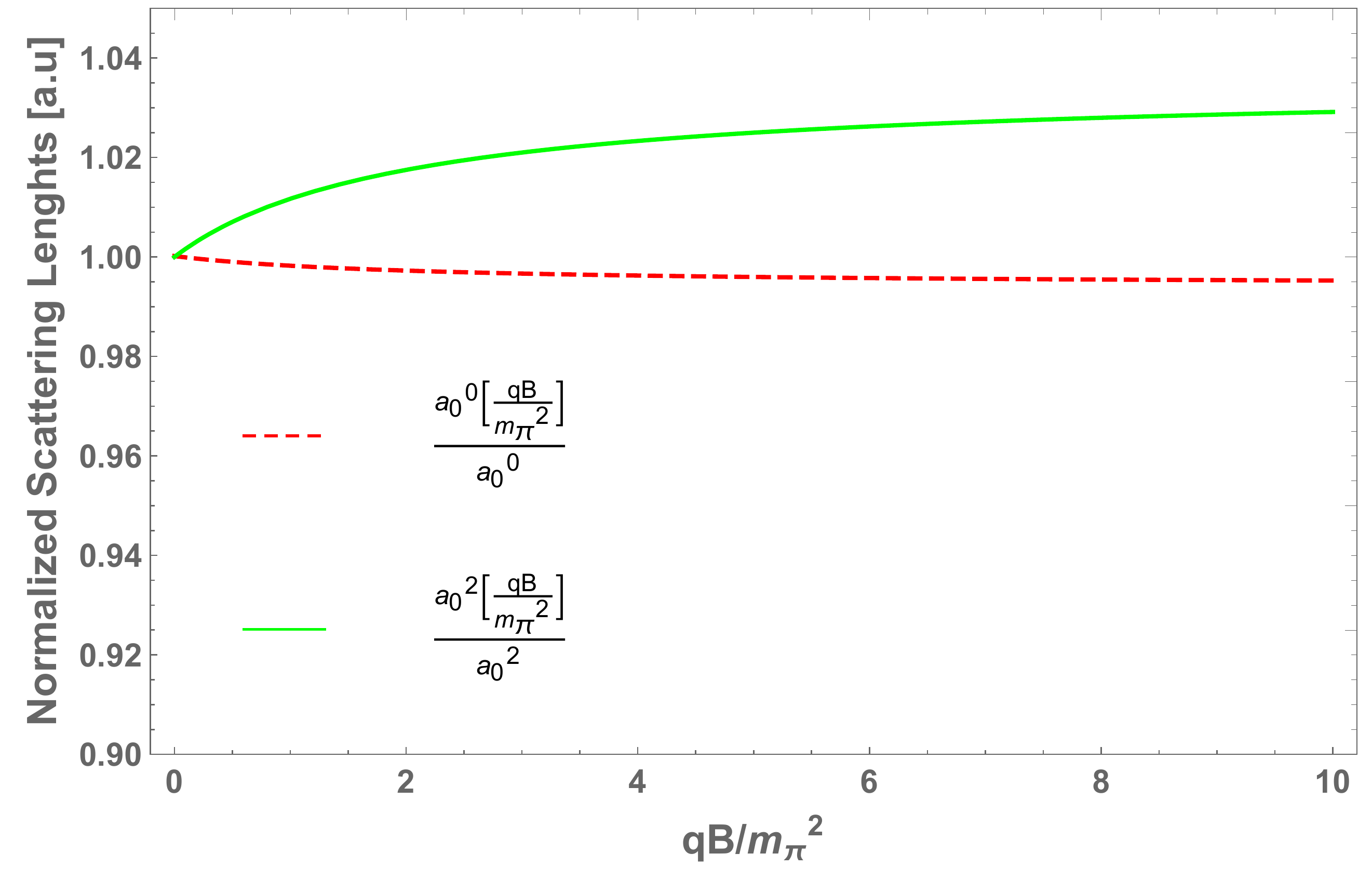}
\caption{(Color online) The scattering parameters $a_0^{0}$ (dashed) and $a_0^{2}$ (solid) are displayed as a function of the dimensionless magnetic field $qB/m_{\pi}^2$. In the figure, the
parameters are normalized by their (experimental) zero-field values.}
\label{as}
\end{figure}

\section{Results and Conclusions}

We have presented a novel method to calculate the scattering lengths for $\pi-\pi$ scattering within the linear sigma model at the one-loop level, in the isospin channels $I = \{0,2\}$, as functions of the external magnetic field intensity. Our calculation shows that the relevant contributions can be reduced to the calculation of two types of ``Fish-type'' diagrams (see
Fig.~\ref{fig:a diagrams}). Along this article, we have obtained exact analytical results for the integrals involved in those diagrams and, moreover, we developed a
regularization procedure that allows to connect smoothly and continuously the low and high magnetic field intensity regimes. Explicit analytical expressions
for the regularized integrals are presented in Eq.~(\ref{eq_I1_Reg}) and Eq.~(\ref{eq_I2_Reg}), respectively. This method extends our previous results~\cite{Leandro} to the full range of magnetic field intensities, thus revealing that
the scattering lengths are smooth and continuous functions of the field (see Fig.~\ref{as}). In particular, our analytical results show that the scattering length $a_0^{0}$ 
decreases  as a function of the magnetic field with respect to its experimental value. On the contrary, the scattering length $a_0^{2}$ is a monotonically increasing function (in absolute value) of the external magnetic field. Interestingly, both scattering lengths achieve asymptotic constant values in the infinitely strong field limit. Remarkably, the observed trends are
opposite to the ones predicted as a function of temperature~\cite{TLSM}, thus suggesting a potential means to experimentally disentangle thermal and magnetic effects. 
As a natural extension of this work, we are currently examining the combined effect of thermal and magnetic contributions of the scattering lengths in these isospin channels. Results shall be reported elsewhere.

\section*{ACKNOWLEDGMENTS}

 M. Loewe acknowledges support from FONDECYT (Chile) under grants No. 1170107, No. 1150471, No. 1150847 and ConicytPIA/BASAL (Chile) grant No. FB0821, L. Monje acknowledges support from FONDECYT (Chile) under grant No. 1170107, AR acknowledges support form ``Consejo Nacional de Ciencia y Tecnolog\'{\i}a (Mexico) under grant, 256494 and R. Zamora would like to thank support from CONICYT FONDECYT Iniciaci\'on under grant No. 11160234.

\appendix*
\section{Integrals over $k_{\parallel} = (k_0,k_3)$}
Here we present in detail the calculation of the integrals involved in Eq.(\ref{eq_I1_reduced}) of the main text. Using the definition of the ``parallel" propagators Eq.(\ref{eq_prop_Land}),
we have
\begin{align}
G_{l,l}(p_0) &= \int\int \frac{d k_0 dk_3}{(2\pi)^2} i\Delta_{l}^{B}(k_0,k_3) i\Delta_{l}^{B}(k_0 - 2p_0,k_3)\nonumber\\ 
&= \frac{i^2}{(2\pi)^2}\int_{-\infty}^{+\infty} dk_3 f_l(k_3,p_0)
\label{eq_ap_prop}
\end{align}
where we have defined the integral
\begin{eqnarray}
f_l(k_3,p_0) = \int_{-\infty}^{+\infty}   \frac{dk_0}{A(k_3)C(k_3)}\;,
\end{eqnarray}
with 
\begin{eqnarray}
A(k_3)&=&k_0^2 - E_l(k_3)^2 + i\epsilon,\nonumber\\
C(k_3)&=&(k_0 - 2 p_0)^2- E_l(k_3)^2 + i\epsilon,
\end{eqnarray}
and $E_l(k_3) = \sqrt{k_3^2 + m_{\pi}^2 + qB(2l + 1)}$.
The integral can be evaluated on the complex $k_0$-plane, by noticing that it possesses four simple
poles at $k_0=\pm E_l(k_3) \mp i\epsilon'$ and $k_0=2p_0 \pm E_l(k_3) \mp i\epsilon'$, i.e., two of them located on the positive imaginary plane, while the other
two are located on the negative imaginary plane. We choose an integration contour as a semicircle, that closes on the upper imaginary plane, and thus
it encloses the poles $k_0^{(1)} = -E_l(k_3) + i\epsilon'$ and $k_0^{(2)} = 2p_0 - E_l(k_3) + i\epsilon'$. By direct application of the residue theorem, we have that 
\begin{eqnarray}
f_l(k_3,p_0) &=& \frac{-2i\pi }{8 p_0}\Bigg[ \frac{1}{E_l(k_3)(E_l(k_3) + p_0)}\nonumber\\
 &&- \frac{1}{E_l(k_3)(E_l(k_3) - p_0)}\Bigg]\;.
\label{eq_f}
\end{eqnarray}
Now we calculate the integral over $k_3$. Inserting Eq.(\ref{eq_f}) into Eq.(\ref{eq_ap_prop}), we have
\begin{eqnarray}
G_{l,l}(p_0) = \frac{i}{16 \pi p_0}\left(g_l(p_0) - g_l(-p_0) \right)\;,
\label{eq_ap_prop2}
\end{eqnarray}
where we have defined
\begin{eqnarray}
g_l(p_0) &=& \int_{-\infty}^{+\infty} \frac{dk_3}{E_l(k_3)[E_l(k_3) - p_0]}\nonumber\\
&=& 2 \int_{0}^{+\infty} \frac{dk_3}{E_l(k_3)[E_l(k_3) - p_0]}\nonumber\\
&=& 2\frac{\frac{\pi}{2} - {\rm{Arctan}}\left(\frac{p_0}{\sqrt{qB(2l+1) + m_{\pi}^2-p_0^2}} \right)}{\sqrt{qB(2l+1) + m_{\pi}^2-p_0^2}}.
\end{eqnarray}
Substituting back into Eq.(\ref{eq_ap_prop2}), we obtain the finite result
\begin{eqnarray}
G_{l,l}(p_0) = \ \frac{i}{4 \pi p_0}\frac{{\rm{Arctan}}\left(\frac{p_0}{\sqrt{qB(2l+1) + m_{\pi}^2-p_0^2}} \right)}{\sqrt{qB(2l+1) + m_{\pi}^2-p_0^2}}.
\end{eqnarray}
Inserting this expression back into Eq.(\ref{eq_I1_reduced}) of the main text, we obtain the infinite series representation
\begin{eqnarray}
I_1[B,p_0] &=&\\
&&\hspace{-5mm} \frac{i qB}{8 \pi^2 p_0}\sum_{l=0}^{\infty}\frac{{\rm{Arctan}}\left(\frac{p_0}{\sqrt{qB(2l+1) + m_{\pi}^2-p_0^2}} \right)}{\sqrt{qB(2l+1) + m_{\pi}^2-p_0^2}}.\nonumber
\end{eqnarray}

\end{document}